\pdfoutput=1
\documentclass[prl,aps,twocolumn,preprintnumbers,amssymb,nobibnotes,nofootinbib,superscriptaddress,raggedbottom,10pt]{revtex4-1}
\usepackage{hyperref,graphicx,array,multirow,color,amsmath,mathrsfs,titlesec,shuffle,tikz}
\usetikzlibrary{calc,decorations.markings}

\titleformat{\section}{\centering\normalsize\normalfont\bf}{\thesection}{0em}{}
\hypersetup{pdftitle={},pdfcreator={},linkcolor=[rgb]{0.15,0.35,0.75},colorlinks=true,citecolor=[rgb]{0.675,0,0.2},urlcolor=[rgb]{0.15,0.35,0.65}}

\newcommand{\fwbox}[2]{\text{\makebox[#1][c]{$\hspace{-150pt}\displaystyle#2\hspace{-150pt}$}}}
\newcommand{\fwboxL}[2]{\text{\makebox[#1][l]{$#2$}}}
\newcommand{\fwboxR}[2]{\text{\makebox[#1][r]{$#2$}}}
\newcommand{\bigger}[1]{\raisebox{-0.95pt}{\scalebox{1.25}{$#1$}}}

\newcommand{\eq}[1]{\vspace{-3.5pt}\begin{equation}\hspace{2pt}#1\hspace{-0pt}\vspace{-3.5pt}\end{equation}}
\newcommand{\mi}{\raisebox{0.75pt}{\scalebox{0.75}{$\hspace{-1pt}\,-\,\hspace{-0.75pt}$}}}
\renewcommand{\pl}{\raisebox{0.75pt}{\scalebox{0.75}{$\hspace{-1pt}\,+\,\hspace{-0.75pt}$}}}
\newcommand{\ab}[1]{\langle #1\rangle}
\newcommand{\N}{\mathcal{N}}
\newcommand{\equivR}{\fwbox{14.5pt}{\hspace{-0pt}\fwboxR{0pt}{\raisebox{0.47pt}{\hspace{1.25pt}:\hspace{-4pt}}}=\fwboxL{0pt}{}}}
\newcommand{\equivL}{\fwbox{14.5pt}{\fwboxR{0pt}{}=\fwboxL{0pt}{\raisebox{0.47pt}{\hspace{-4pt}:\hspace{1.25pt}}}}}
\newcommand{\ddmColor}[3]{f^{\fwbox{10pt}{{\color{ord}\vec{#2}}}}_{\fwbox{10pt}{{\color{anchorLeg}#1}\hspace{2.pt}{\color{anchorLeg}#3}}}}
\definecolor{hblue}{rgb}{0,0,0.575}
\definecolor{hred}{rgb}{0.575,0.0,0.225}
\definecolor{hteal}{rgb}{0.0,0.545,0.7451}
\definecolor{mhvblue}{rgb}{0.6,0.6,0.7765}
\definecolor{ampgrey}{rgb}{0.9,0.9,0.9}
\definecolor{unord}{rgb}{0,0,0}
\definecolor{ord}{rgb}{0,0,0.575}
\definecolor{anchorLeg}{rgb}{0.575,0.0,0.225}

\def\figScale{0.825}\def\edgeLen{1*\figScale}\def\fScale{\footnotesize}\pgfmathsetmacro{\pLen}{\edgeLen/(2*sin(72/2))}\def\legLen{\edgeLen*0.55}\def\dotDist{\legLen*0.75}\def\labelDist{\legLen*1.4}\def\lineThickness{(1pt)}\def\dotSize{(\figScale*1pt)}\def\ampSize{(1*\figScale*12pt)}\def\eph{0.4}
\tikzset{fullamp/.style={coordinate,minimum size=1.5*\ampSize,ball color=black!20,circle,text=white,inner sep=0}}\tikzset{fullmhv/.style={coordinate,minimum size=0.9*\ampSize,ball color=mhvblue,circle,text=white,inner sep=0}}\tikzset{fullmhvBar/.style={coordinate,minimum size=0.9*\ampSize,ball color=white,circle,text=white,inner sep=0}}\tikzset{ordAmp/.style={fill=ampgrey,circle,draw=black,line width=\lineThickness,minimum size=1.5*0.8*\ampSize,text=white,inner sep=0}}\tikzset{mhv/.style={fill=mhvblue,circle,draw=black,line width=\lineThickness,minimum size=0.8*\ampSize,text=white,inner sep=0}}\tikzset{mhvBar/.style={fill=white,circle,draw=black,line width=\lineThickness,minimum size=0.8*\ampSize,text=white,inner sep=0}}\tikzset{fgraphEdge/.style={anchorLeg,line width=\lineThickness,line cap=round}}\tikzset{fgraphExt/.style={ord,line width=\lineThickness,line cap=round}}\tikzset{fgraphOpt/.style={ord,dotted,line width=\lineThickness,line cap=round}}\tikzset{fdot/.style={fill=anchorLeg,circle,minimum size=0.35*\ampSize,inner sep=0}}\tikzset{bdot/.style={fill=black,circle,minimum size=0.35*\ampSize,inner sep=0}}\tikzset{ext/.style={black,line width=\lineThickness,line cap=round}}\tikzset{under/.style={white,line width=4*\lineThickness,line cap=round}}\tikzset{optExt/.style={black,dotted,line width=\lineThickness,line cap=round}}\tikzset{ddot/.style={circle,black,draw,radius=\dotSize,fill=black,inner sep=0pt}}
\newcommand{\ddmFgraphExpanded}{\begin{tikzpicture}[baseline=-3.05]\coordinate (f1) at (0,0);\coordinate (f2) at ($(f1)+1.25*\eph*(1,0)$);\coordinate (f3) at ($(f1)+3.5*\eph*(1,0)$);\coordinate (f4) at ($(f1)+4.75*\eph*(1,0)$);\coordinate (e0) at ($(f1)+(-135:\eph)$);\coordinate (e1) at ($(f1)+(90:0.9*\eph)$);\coordinate (e2) at ($(f2)+(90:0.9*\eph)$);\coordinate (e2) at ($(f2)+(90:0.9*\eph)$);\coordinate (e3) at ($(f3)+(90:0.9*\eph)$);\coordinate (e4) at ($(f4)+(90:0.9*\eph)$);\coordinate (en) at ($(f4)+(-45:\eph)$);\draw[fgraphEdge](e0)--(f1)--(f2)--(f3)--(f4)--(en);\draw[fgraphExt](f1)--(e1);\draw[fgraphExt](f2)--(e2);\draw[fgraphExt](f3)--(e3);\draw[fgraphExt](f4)--(e4);\foreach \step in {0.3,0.5,0.7} {\node at ($(f2)!\step!(f3)+(90:0.5*\eph)$) [ddot,ord]{$$};};\node[fdot] at (f1) {};\node[fdot] at (f2) {};\node[fdot] at (f3) {};\node[fdot] at (f4) {};\node[above=2pt,anchor=north] at (e0) {{\footnotesize{\color{anchorLeg}$\alpha\,\,$}}};\node[above=2pt,anchor=base] at (e1) {{\footnotesize{\color{ord}$\,\,a_1$}}};\node[above=2pt,anchor=base] at (e2) {{\footnotesize{\color{ord}$\,\,a_2$}}};\node[above=2pt,anchor=base] at (e3) {{\footnotesize{\color{ord}$\,\,\,a_{\text{-}2}$}}};\node[above=2pt,anchor=base] at (e4) {{\footnotesize{\color{ord}$\,\,\,a_{\text{-}1}$}}};\node[above=2pt,anchor=north] at (en) {{\footnotesize{\color{anchorLeg}$\,\,\beta$}}};\end{tikzpicture}}
\newcommand{\ddmFgraphContracted}{\begin{tikzpicture}[baseline=-3.05]\coordinate (f1) at (0,0);\coordinate (e0) at ($(f1)+(-135:\eph)$);\coordinate (en) at ($(f1)+(-45:\eph)$);\coordinate (e1) at ($(f1)+(90:0.9*\eph)$);\draw[fgraphEdge](e0)--(f1)--(en);\node[above=2pt,anchor=north] at (e0) {{\footnotesize{\color{anchorLeg}$\alpha\,\,$}}};\node[above=2pt,anchor=north] at (en) {{\footnotesize{\color{anchorLeg}$\,\,\beta$}}};\foreach \step in {-39,39} {\draw[fgraphOpt] (f1)-- ($(f1)+(90+\step:\legLen)$);};\foreach \step in {-13,13} {\node at ($(f1)+(90+\step:0.8\dotDist)$) [ddot,ord]{$$};};\node[fdot] at (f1) {};\node [above=2pt,anchor=base]at (e1)  {{\footnotesize{\color{ord}$\vec{a}$}}};\end{tikzpicture}}
\newcommand{\egHexaboxFull}{\begin{tikzpicture}[scale=\figScale,baseline=-3.05]\useasboundingbox ($\figScale*(-1.75,-1.75)$) rectangle ($\figScale*(2.75,1.75)$);\coordinate (v1) at ($(-90:0.8*\pLen)+\figScale*(1.05,0)$);\coordinate (v3) at ($(-3*72+36:\pLen)+(0.275,0)$);\coordinate (v5) at ($(90:0.8*\pLen)+\figScale*(1.05,0)$);\coordinate (v6) at ($(0:0.8*\pLen)+\figScale*(1.05,0)-\figScale*(0.225,0)$);\coordinate (v7) at ($(180:0.8*\pLen)+\figScale*(1.05,0)+\figScale*(0.225,0)$);\coordinate (v4) at ($(v5)-\figScale*(1.25,0)$);\coordinate (v2) at ($(v1)-\figScale*(1.25,0)$);\draw[ext] (v1)--(v2)--(v3)--(v4)--(v5)--(v6)--(v1)--(v7)--(v5);\foreach \step in {-30,30} {\draw[optExt] (v1)-- ($(v1)+(-65+\step:\legLen)$);};\foreach \step in {-10,10} {\node at ($(v1)+(-65+\step:\dotDist)$) [ddot]{$$};};\node at ($(v1)+(-65:\labelDist)$) {{\fScale$C$}};\node at (v1) [fullmhv] {};\draw[ext] (v2)--($(v2)+(-36-72:\legLen)$);\node at ($(v2)+(-36-72:\labelDist)$) {{\fScale$\alpha$}};\node at (v2) [fullmhvBar] {};\foreach \step in {-39,39} {\draw[optExt] (v3)-- ($(v3)+(180+\step:\legLen)$);};\foreach \step in {-13,13} {\node at ($(v3)+(180+\step:\dotDist)$) [ddot]{$$};};\node at ($(v3)+(0.1,0)+(-36-2*72:\labelDist)$) {{\fScale$A$}};\node at (v3) [fullmhv] {};\draw[ext] (v4)--($(v4)+(-36-3*72:\legLen)$);\node at ($(v4)+(-36-3*72:\labelDist)$) {{\fScale$\beta$}};\node at (v4) [fullmhvBar] {};\foreach \step in {-27,27} {\draw[optExt] (v5)-- ($(v5)+(65+\step:\legLen)$);};\foreach \step in {-9,9} {\node at ($(v5)+(65+\step:\dotDist)$) [ddot]{$$};};\node[] at ($(v5)+(65:\labelDist)$) {{\fScale$\hspace{-0pt}B$}};\node at (v5) [fullmhv] {};\draw[ext] (v6)--($(v6)+(0:\legLen)$);\node at ($(v6)+(0:\labelDist)$) {{\fScale$\gamma\,\,$}};\node at (v6) [fullmhvBar] {};\draw[ext] (v7)--($(v7)+(180:\legLen)$);\node at ($(v7)+(180:\labelDist)$) {{\fScale$\,\,\delta$}};\node at (v7) [fullmhvBar] {};\node at (v6) [fullmhvBar] {};\end{tikzpicture}}
\newcommand{\egHexaboxOrdered}{\begin{tikzpicture}[scale=\figScale,baseline=-3.05]\useasboundingbox ($\figScale*(-1.75,-1.75)$) rectangle ($\figScale*(2.75,1.75)$);\coordinate (v1) at ($(-90:0.8*\pLen)+\figScale*(1.05,0)$);\coordinate (v3) at ($(-3*72+36:\pLen)+(0.275,0)$);\coordinate (v5) at ($(90:0.8*\pLen)+\figScale*(1.05,0)$);\coordinate (v6) at ($(0:0.8*\pLen)+\figScale*(1.05,0)-\figScale*(0.225,0)$);\coordinate (v7) at ($(180:0.8*\pLen)+\figScale*(1.05,0)+\figScale*(0.225,0)$);\coordinate (v4) at ($(v5)-\figScale*(1.25,0)$);\coordinate (v2) at ($(v1)-\figScale*(1.25,0)$);\draw[ext] (v1)--(v2)--(v3)--(v4)--(v5)--(v6)--(v1)--(v7)--(v5);\foreach \step in {-30,30} {\draw[optExt] (v1)-- ($(v1)+(-65+\step:\legLen)$);};\foreach \step in {-10,10} {\node at ($(v1)+(-65+\step:\dotDist)$) [ddot]{$$};};\node at ($(v1)+(-65:\labelDist)$) {{\fScale$\vec{c}_1$}};\foreach \step in {-10,10} {\draw[optExt] (v1)-- ($(v1)+(155+\step:\legLen*1.4)$);};\foreach \step in {0} {\node at ($(v1)+(155+\step:\dotDist*1.4)$) [ddot]{$$};};\node at ($(v1)+(155:\labelDist*1.35)+(0.075,0)$) {{\fScale$\vec{c}_2$}};\node at (v1) [mhv] {};\draw[ext] (v2)--($(v2)+(-36-72:\legLen)$);\node at ($(v2)+(-36-72:\labelDist)$) {{\fScale$\alpha$}};\node at (v2) [mhvBar] {};\foreach \step in {-39,39} {\draw[optExt] (v3)-- ($(v3)+(180+\step:\legLen)$);};\foreach \step in {-13,13} {\node at ($(v3)+(180+\step:\dotDist)$) [ddot]{$$};};\node at ($(v3)+(-36-2*72:0.9*\labelDist)$) {{\fScale$\vec{a}\!$}};\node at (v3) [mhv] {};\draw[ext] (v4)--($(v4)+(-36-3*72:\legLen)$);\node at ($(v4)+(-36-3*72:\labelDist)$) {{\fScale$\beta$}};\node at (v4) [mhvBar] {};\foreach \step in {-27,27} {\draw[optExt] (v5)-- ($(v5)+(65+\step:\legLen)$);};\foreach \step in {-9,9} {\node at ($(v5)+(65+\step:\dotDist)$) [ddot]{$$};};\node[] at ($(v5)+(65:\labelDist)$) {{\fScale$\hspace{-0pt}\vec{b}_1$}};\foreach \step in {-10,10} {\draw[optExt] (v5)-- ($(v5)+(-155+\step:\legLen*1.4)$);};\foreach \step in {0} {\node at ($(v5)+(-155+\step:\dotDist*1.4)$) [ddot]{$$};};\node[] at ($(v5)+(-160:\labelDist*1.25)+(0.075,0)$) {{\fScale$\hspace{-0pt}\vec{b}_2$}};\node at (v5) [mhv] {};\draw[ext] (v6)--($(v6)+(0:\legLen)$);\node at ($(v6)+(0:\labelDist)$) {{\fScale$\gamma\,\,$}};\node at (v6) [mhvBar] {};\draw[ext] (v7)--($(v7)+(180:\legLen)$);\node at ($(v7)+(180:\labelDist)$) {{\fScale$\,\,\delta$}};\node at (v7) [mhvBar] {};\node at (v6) [mhvBar] {};\end{tikzpicture}}
\newcommand{\egHexaboxColour}{\begin{tikzpicture}[scale=\figScale,baseline=-3.05]\useasboundingbox ($\figScale*(-1.75,-1.75)$) rectangle ($\figScale*(2.75,1.75)$);\coordinate (v1) at ($(-90:0.8*\pLen)+\figScale*(1.05,0)$);\coordinate (v3) at ($(-3*72+36:\pLen)+(0.275,0)$);\coordinate (v5) at ($(90:0.8*\pLen)+\figScale*(1.05,0)$);\coordinate (v6) at ($(0:0.8*\pLen)+\figScale*(1.05,0)-\figScale*(0.225,0)$);\coordinate (v7) at ($(180:0.8*\pLen)+\figScale*(1.05,0)+\figScale*(0.225,0)$);\coordinate (v4) at ($(v5)-\figScale*(1.25,0)$);\coordinate (v2) at ($(v1)-\figScale*(1.25,0)$);\coordinate (v5a) at ($(v5)!0.3!(v6)$);\coordinate (v5b) at ($(v5)!0.3!(v7)$);\coordinate (v1a) at ($(v1)!0.3!(v6)$);\coordinate (v1b) at ($(v1)!0.3!(v7)$);\draw[fgraphEdge] (v1)--(v2)--(v3)--(v4)--(v5)--(v6)--(v1)--(v7)--(v5);\foreach \step in {-30,30} {\draw[fgraphOpt] (v1a)-- ($(v1a)+(-45+\step:\legLen*.75)$);};\foreach \step in {-10,10} {\node at ($(v1a)+(-45+\step:\dotDist*.75)$) [ddot,ord]{$$};};\node at ($(v1a)+(-45:\labelDist*.85)$) {{\fScale${\color{ord}\vec{c}_1}$}};\foreach \step in {-15,15} {\draw[fgraphOpt] (v1b)-- ($(v1b)+(175+\step:\legLen*1)$);};\foreach \step in {0} {\node at ($(v1b)+(175+\step:\dotDist*1)$) [ddot,ord]{$$};};\node at ($(v1)+(165:\labelDist*1.35)+(0.125,0)$) {{\fScale${\color{ord}\vec{c}_2}$}};\node at (v1) [fdot] {};\node at (v1a) [fdot] {};\node at (v1b) [fdot] {};\draw[fgraphEdge] (v2)--($(v2)+(-36-72:\legLen*.75)$);\node at ($(v2)+(-36-72:\labelDist*.75)$) {{\fScale${\color{anchorLeg}\alpha}$}};\node at (v2) [fdot] {};\foreach \step in {-39,39} {\draw[fgraphOpt] (v3)-- ($(v3)+(180+\step:\legLen*.75)$);};\foreach \step in {-13,13} {\node at ($(v3)+(180+\step:\dotDist*.75)$) [ddot,ord]{$$};};\node at ($(v3)+(-36-2*72:0.9*\labelDist*.75)$) {{\fScale${\color{ord}\vec{a}}$}};\node at (v3) [fdot] {};\draw[fgraphEdge] (v4)--($(v4)+(-36-3*72:\legLen*.75)$);\node at ($(v4)+(-36-3*72:\labelDist*.75)$) {{\fScale${\color{anchorLeg}\beta}$}};\node at (v4) [fdot] {};\foreach \step in {-27,27} {\draw[fgraphOpt] (v5a)-- ($(v5a)+(45+\step:\legLen*.75)$);};\foreach \step in {-9,9} {\node at ($(v5a)+(45+\step:\dotDist*.75)$) [ddot,ord]{$$};};\node[] at ($(v5a)+(45:\labelDist*.85)$) {{\fScale$\hspace{-0pt}{\color{ord}\vec{b}_1}$}};\foreach \step in {-15,15} {\draw[fgraphOpt] (v5b)-- ($(v5b)+(-172+\step:\legLen*1.)$);};\foreach \step in {0} {\node at ($(v5b)+(-172+\step:\dotDist*1.)$) [ddot,ord]{$$};};\node[] at ($(v5)+(-165:\labelDist*1.35)+(0.125,0)$) {{\fScale$\hspace{-0pt}{\color{ord}\vec{b}_2}$}};\node at (v5) [fdot] {};\node at (v5a) [fdot] {};\node at (v5b) [fdot] {};\draw[fgraphEdge] (v6)--($(v6)+(0:\legLen*.75)$);\node at ($(v6)+(0:\labelDist*.75)$) {{\fScale${\color{anchorLeg}\gamma}\,\,$}};\node at (v6) [fdot] {};\draw[fgraphEdge] (v7)--($(v7)+(180:\legLen)$);\node at ($(v7)+(180:\labelDist)$) {{\fScale${\color{anchorLeg}\,\,\delta}$}};\node at (v7) [fdot] {};\node at (v6) [fdot] {};\end{tikzpicture}}
\newcommand{\egHeptacut}{\begin{tikzpicture}[scale=\figScale,baseline=-3.05]\useasboundingbox ($\figScale*(-1.75,-1.75)$) rectangle ($\figScale*(2.75,1.75)$);\coordinate (v1) at ($(-90:0.8*\pLen)+\figScale*(1.05,0)$);\coordinate (v3) at ($(-3*72+36:\pLen)+(0.275,0)$);\coordinate (v5) at ($(90:0.8*\pLen)+\figScale*(1.05,0)$);\coordinate (v6) at ($(0:0.8*\pLen)+\figScale*(1.05,0)-\figScale*(0.125,0)$);\coordinate (v7) at ($(180:0.8*\pLen)+\figScale*(1.05,0)+\figScale*(0.225,0)$);\coordinate (v4) at ($(v5)-\figScale*(1.45,0)$);\coordinate (v2) at ($(v1)-\figScale*(1.45,0)$);\draw[ext] (v1)--(v2)--(v4)--(v5)--(v6)--(v1)--(v7)--(v5);\draw[ext,decorate,decoration={markings,mark=at position 0.525cm with {\arrow{stealth}}}] (v2)--(v4);\draw[ext,decorate,decoration={markings,mark=at position 0.365cm with {\arrow{stealth}}}] (v5)--(v6);\node[anchor=east] at ($(v2)!0.5!(v4)$) {{$\;\;\ell_1\!$}};\node[anchor=west] at ($(v5)!0.25!(v6)$) {{$\,\ell_2$}};\foreach \step in {-30,30} {\draw[optExt] (v1)-- ($(v1)+(-65+\step:\legLen)$);};\foreach \step in {-10,10} {\node at ($(v1)+(-65+\step:\dotDist)$) [ddot]{$$};};\node at ($(v1)+(-65:\labelDist)$) {{$B$}};\node at (v1) [fullmhv] {};\draw[ext] (v2)--($(v2)+(-135:\legLen)$);\node at ($(v2)+(-135:\labelDist)$) {{$1$}};\node at (v2) [fullmhv] {};\draw[ext] (v4)--($(v4)+(135:\legLen)$);\node at ($(v4)+(135:\labelDist)$) {{$2$}};\node at (v4) [fullmhvBar] {};\foreach \step in {-27,27} {\draw[optExt] (v5)-- ($(v5)+(65+\step:\legLen)$);};\foreach \step in {-9,9} {\node at ($(v5)+(65+\step:\dotDist)$) [ddot]{$$};};\node[] at ($(v5)+(65:\labelDist)$) {{$\hspace{-0pt}A$}};\node at (v5) [fullmhv] {};\draw[ext] (v6)--($(v6)+(0:\legLen)$);\node at ($(v6)+(0:\labelDist)$) {{$4\,\,$}};\node at (v6) [fullmhvBar] {};\draw[ext] (v7)--($(v7)+(180:\legLen)$);\node at ($(v7)+(180:\labelDist)$) {{$\,\,3$}};\node at (v7) [fullmhvBar] {};\node at (v6) [fullmhvBar] {};\end{tikzpicture}}
\newcommand{\egGRTseed}{\begin{tikzpicture}[scale=\figScale,baseline=-3.05]\useasboundingbox ($\figScale*(-1.5,-1.5)$) rectangle ($\figScale*(2.5,1.5)$);\coordinate (v1) at ($(-90:0.8*\pLen)+\figScale*(1.05,0)$);\coordinate (v3) at ($(-3*72+36:\pLen)+(0.275,0)$);\coordinate (v5) at ($(90:0.8*\pLen)+\figScale*(1.05,0)$);\coordinate (v6) at ($(0:0.8*\pLen)+\figScale*(1.05,0)-\figScale*(0.125,0)$);\coordinate (v7) at ($(180:0.8*\pLen)+\figScale*(1.05,0)+\figScale*(0.125,0)$);\coordinate (v4) at ($(v5)-\figScale*(1.55,0)$);\coordinate (v2) at ($(v1)-\figScale*(1.55,0)$);\draw[ext] (v1)--(v2)--(v4)--(v5)--(v6)--(v1)--(v7)--(v5);\foreach \step in {-30,30} {\draw[optExt] (v1)-- ($(v1)+(-65+\step:\legLen)$);};\foreach \step in {-10,10} {\node at ($(v1)+(-65+\step:\dotDist)$) [ddot]{$$};};\node at ($(v1)+(-65:\labelDist)$) {{$C$}};\node at (v1) [fullmhv] {};\foreach \step in {-39} {\draw[optExt] (v2)-- ($(v2)+(-135+\step:\legLen)$);};\foreach \step in {39} {\draw[ext] (v2)-- ($(v2)+(-135+\step:\legLen)$);};\foreach \step in {-13,13} {\node at ($(v2)+(-135+\step:\dotDist)$) [ddot]{$$};};\node at ($(v2)+(0.0,0)+(-135:\labelDist)$) {{$A$}};\node at (v2) [fullmhv] {};\draw[ext] (v4)--($(v4)+(135:\legLen)$);\node at ($(v4)+(135:\labelDist)$) {{$\alpha$}};\node at (v4) [fullmhvBar] {};\foreach \step in {-27,27} {\draw[optExt] (v5)-- ($(v5)+(65+\step:\legLen)$);};\foreach \step in {-9,9} {\node at ($(v5)+(65+\step:\dotDist)$) [ddot]{$$};};\node[] at ($(v5)+(65:\labelDist)$) {{$\hspace{-0pt}B$}};\node at (v5) [fullmhv] {};\draw[ext] (v6)--($(v6)+(0:\legLen)$);\node at ($(v6)+(0:\labelDist)$) {{$\beta\,\,$}};\node at (v6) [fullmhvBar] {};\draw[ext] (v7)--($(v7)+(180:\legLen)$);\node at ($(v7)+(180:\labelDist)$) {{$\,\,\gamma$}};\node at (v7) [fullmhvBar] {};\node at (v6) [fullmhvBar] {};\end{tikzpicture}}
\newcommand{\egGRTtermA}{\begin{tikzpicture}[scale=\figScale,baseline=-3.05]\useasboundingbox ($\figScale*(-0.65,-1.5)$) rectangle ($\figScale*(2.8,1.5)$);\coordinate (v1) at ($(-90:0.8*\pLen)+\figScale*(1.05,0)$);\coordinate (v3) at ($(-3*72+36:\pLen)+(0.275,0)$);\coordinate (v5) at ($(90:0.8*\pLen)+\figScale*(1.05,0)$);\coordinate (v7) at ($(v1)!0.5!(v5)$);\coordinate (v4) at ($(v5)-\figScale*(0.95,0)$);\coordinate (v2) at ($(v1)-\figScale*(0.95,0)$);\coordinate (v6) at ($(v5)+\figScale*(0.95,0)$);\coordinate (v8) at ($(v1)+\figScale*(0.95,0)$);\draw[ext] (v1)--(v2)--(v4)--(v5)--(v6)--(v8)--(v1)--(v7)--(v5);\draw[ext,hteal,line width=2pt] (v1)--(v8);\foreach \step in {-39,39} {\draw[optExt] (v8)-- ($(v8)+(-45+\step:\legLen)$);};\foreach \step in {-13,13} {\node at ($(v8)+(-45+\step:\dotDist)$) [ddot]{$$};};\node at ($(v8)+(-45:\labelDist)$) {{$C$}};\node at (v8) [fullmhv] {};\foreach \step in {-39} {\draw[optExt] (v2)-- ($(v2)+(-135+\step:\legLen)$);};\foreach \step in {39} {\draw[ext] (v2)-- ($(v2)+(-135+\step:\legLen)$);};\foreach \step in {-13,13} {\node at ($(v2)+(-135+\step:\dotDist)$) [ddot]{$$};};\node at ($(v2)+(0.0,0)+(-135:\labelDist)$) {{$A$}};\node at (v2) [fullmhv] {};\draw[ext] (v4)--($(v4)+(135:\legLen)$);\node at ($(v4)+(135:\labelDist)$) {{$\alpha$}};\node at (v4) [fullmhvBar] {};\foreach \step in {-27,27} {\draw[optExt] (v5)-- ($(v5)+(90+\step:\legLen)$);};\foreach \step in {-9,9} {\node at ($(v5)+(90+\step:\dotDist)$) [ddot]{$$};};\node[] at ($(v5)+(90:\labelDist)$) {{$\hspace{-0pt}B$}};\node at (v5) [fullmhv] {};\draw[ext] (v6)--($(v6)+(45:\legLen)$);\node at ($(v6)+(45:\labelDist)$) {{$\,\,\beta$}};\node at (v6) [fullmhvBar] {};\draw[ext] (v7)--($(v7)+(180:\legLen)$);\node at ($(v7)+(180:\labelDist)$) {{$\,\,\gamma$}};\node at (v7) [fullmhvBar] {};\node at (v1) [fullmhvBar] {};\node at (v6) [fullmhvBar] {};\end{tikzpicture}}
\newcommand{\egGRTtermB}{\begin{tikzpicture}[scale=\figScale,baseline=-3.05]\useasboundingbox ($\figScale*(-0.65,-1.5)$) rectangle ($\figScale*(2.8,1.5)$);\coordinate (v1) at ($(-90:0.8*\pLen)+\figScale*(1.05,0)$);\coordinate (v3) at ($(-3*72+36:\pLen)+(0.275,0)$);\coordinate (v5) at ($(90:0.8*\pLen)+\figScale*(1.05,0)$);\coordinate (v7) at ($(v1)!0.5!(v5)$);\coordinate (v4) at ($(v5)-\figScale*(0.95,0)$);\coordinate (v2) at ($(v1)-\figScale*(0.95,0)$);\coordinate (v6) at ($(v5)+\figScale*(0.95,0)$);\coordinate (v8) at ($(v1)+\figScale*(0.95,0)$);\draw[ext] (v1)--(v2)--(v4)--(v5)--(v6)--(v8)--(v1)--(v7)--(v5);\draw[ext,hteal,line width=2pt] (v1)--(v8);\foreach \step in {-39,39} {\draw[optExt] (v8)-- ($(v8)+(-45+\step:\legLen)$);};\foreach \step in {-13,13} {\node at ($(v8)+(-45+\step:\dotDist)$) [ddot]{$$};};\node at ($(v8)+(-45:\labelDist)$) {{$C$}};\node at (v8) [fullmhv] {};\foreach \step in {-39} {\draw[optExt] (v2)-- ($(v2)+(-135+\step:\legLen)$);};\foreach \step in {39} {\draw[ext] (v2)-- ($(v2)+(-135+\step:\legLen)$);};\foreach \step in {-13,13} {\node at ($(v2)+(-135+\step:\dotDist)$) [ddot]{$$};};\node at ($(v2)+(0.0,0)+(-135:\labelDist)$) {{$A$}};\node at (v2) [fullmhv] {};\draw[ext] (v4)--($(v4)+(135:\legLen)$);\node at ($(v4)+(135:\labelDist)$) {{$\alpha$}};\node at (v4) [fullmhvBar] {};\foreach \step in {-27,27} {\draw[optExt] (v5)-- ($(v5)+(90+\step:\legLen)$);};\foreach \step in {-9,9} {\node at ($(v5)+(90+\step:\dotDist)$) [ddot]{$$};};\node[] at ($(v5)+(90:\labelDist)$) {{$\hspace{-0pt}B$}};\node at (v5) [fullmhv] {};\draw[ext] (v6)--($(v6)+(45:\legLen)$);\node at ($(v6)+(45:\labelDist)$) {{$\,\gamma$}};\node at (v6) [fullmhvBar] {};\draw[ext] (v7)--($(v7)+(180:\legLen)$);\node at ($(v7)+(180:\labelDist)$) {{$\,\,\beta\,$}};\node at (v7) [fullmhvBar] {};\node at (v1) [fullmhvBar] {};\node at (v6) [fullmhvBar] {};\end{tikzpicture}}
\newcommand{\egGRTtermC}{\begin{tikzpicture}[scale=\figScale,baseline=-3.05]\useasboundingbox ($\figScale*(-1.35,-1.5)$) rectangle ($\figScale*(2.4,1.5)$);\coordinate (v1) at ($(-90:0.8*\pLen)+\figScale*(1.05,0)$);\coordinate (v3) at ($(-3*72+36:\pLen)+(0.275,0)$);\coordinate (v5) at ($(90:0.8*\pLen)+\figScale*(1.05,0)$);\coordinate (v6) at ($(0:0.8*\pLen)+\figScale*(1.05,0)-\figScale*(0.225,0)$);\coordinate (v7) at ($(180:0.8*\pLen)+\figScale*(1.05,0)+\figScale*(0.225,0)$);\coordinate (v4) at ($(v5)-\figScale*(1.05,0)$);\coordinate (v2) at ($(v1)-\figScale*(1.05,0)$);\draw[ext] (v1)--(v2)--(v3)--(v4)--(v5)--(v6)--(v1)--(v7)--(v5);\draw[ext,hteal,line width=2pt] (v1)--(v2);\foreach \step in {-30,30} {\draw[optExt] (v1)-- ($(v1)+(-65+\step:\legLen)$);};\foreach \step in {-10,10} {\node at ($(v1)+(-65+\step:\dotDist)$) [ddot]{$$};};\node at ($(v1)+(-65:\labelDist)$) {{$C'$}};\node at (v1) [fullmhv] {};\draw[ext] (v2)--($(v2)+(-36-72:\legLen)$);\node at ($(v2)+(-36-72:\labelDist)$) {{$\!\!\!\!\!\fwboxL{0pt}{c\!\hspace{-1pt}\in\!\!C}$}};\node at (v2) [fullmhvBar] {};\foreach \step in {-39} {\draw[optExt] (v3)-- ($(v3)+(180+\step:\legLen)$);};\foreach \step in {39} {\draw[ext] (v3)-- ($(v3)+(180+\step:\legLen)$);};\foreach \step in {-13,13} {\node at ($(v3)+(180+\step:\dotDist)$) [ddot]{$$};};\node at ($(v3)+(0.1,0)+(-36-2*72:\labelDist)$) {{$A$}};\node at (v3) [fullmhv] {};\draw[ext] (v4)--($(v4)+(-36-3*72:\legLen)$);\node at ($(v4)+(-36-3*72:\labelDist)$) {{$\alpha$}};\node at (v4) [fullmhvBar] {};\foreach \step in {-27,27} {\draw[optExt] (v5)-- ($(v5)+(65+\step:\legLen)$);};\foreach \step in {-9,9} {\node at ($(v5)+(65+\step:\dotDist)$) [ddot]{$$};};\node[] at ($(v5)+(65:\labelDist)$) {{$\hspace{-0pt}B$}};\node at (v5) [fullmhv] {};\draw[ext] (v6)--($(v6)+(0:\legLen)$);\node at ($(v6)+(0:\labelDist)$) {{$\beta\,\,$}};\node at (v6) [fullmhvBar] {};\draw[ext] (v7)--($(v7)+(180:\legLen)$);\node at ($(v7)+(180:\labelDist)$) {{$\,\,\gamma$}};\node at (v7) [fullmhvBar] {};\node at (v6) [fullmhvBar] {};\end{tikzpicture}}
\newcommand{\egGRTtermD}{\begin{tikzpicture}[scale=\figScale,baseline=-3.05]\useasboundingbox ($\figScale*(-1.35,-1.5)$) rectangle ($\figScale*(2.4,1.5)$);\coordinate (v1) at ($(-90:0.8*\pLen)+\figScale*(1.05,0)$);\coordinate (v3) at ($(-3*72+36:\pLen)+(0.275,0)$);\coordinate (v5) at ($(90:0.8*\pLen)+\figScale*(1.05,0)$);\coordinate (v6) at ($(0:0.8*\pLen)+\figScale*(1.05,0)-\figScale*(0.225,0)$);\coordinate (v7) at ($(180:0.8*\pLen)+\figScale*(1.05,0)+\figScale*(0.225,0)$);\coordinate (v4) at ($(v5)-\figScale*(1.05,0)$);\coordinate (v2) at ($(v1)-\figScale*(1.05,0)$);\draw[ext] (v1)--(v2)--(v3)--(v4)--(v5)--(v6)--(v1)--(v7)--(v5);\draw[ext,hteal,line width=2pt] (v2)--(v3);\foreach \step in {-30,30} {\draw[optExt] (v1)-- ($(v1)+(-65+\step:\legLen)$);};\foreach \step in {-10,10} {\node at ($(v1)+(-65+\step:\dotDist)$) [ddot]{$$};};\node at ($(v1)+(-65:\labelDist)$) {{$C$}};\node at (v1) [fullmhv] {};\draw[ext] (v2)--($(v2)+(-36-72:\legLen)$);\node at ($(v2)+(-36-72:\labelDist)$) {{$\!\!\!\!\!\fwboxL{0pt}{a\hspace{-1pt}\!\in\!\!A}$}};\node at (v2) [fullmhvBar] {};\foreach \step in {-39,39} {\draw[optExt] (v3)-- ($(v3)+(180+\step:\legLen)$);};\foreach \step in {-13,13} {\node at ($(v3)+(180+\step:\dotDist)$) [ddot]{$$};};\node at ($(v3)+(0.1,0)+(-36-2*72:\labelDist)$) {{$A'\,$}};\node at (v3) [fullmhv] {};\draw[ext] (v4)--($(v4)+(-36-3*72:\legLen)$);\node at ($(v4)+(-36-3*72:\labelDist)$) {{$\alpha$}};\node at (v4) [fullmhvBar] {};\foreach \step in {-27,27} {\draw[optExt] (v5)-- ($(v5)+(65+\step:\legLen)$);};\foreach \step in {-9,9} {\node at ($(v5)+(65+\step:\dotDist)$) [ddot]{$$};};\node[] at ($(v5)+(65:\labelDist)$) {{$\hspace{-0pt}B$}};\node at (v5) [fullmhv] {};\draw[ext] (v6)--($(v6)+(0:\legLen)$);\node at ($(v6)+(0:\labelDist)$) {{$\beta\,\,$}};\node at (v6) [fullmhvBar] {};\draw[ext] (v7)--($(v7)+(180:\legLen)$);\node at ($(v7)+(180:\labelDist)$) {{$\,\,\gamma$}};\node at (v7) [fullmhvBar] {};\node at (v6) [fullmhvBar] {};\end{tikzpicture}}
\newcommand{\schematicHeptacut}{\begin{tikzpicture}[scale=\figScale,baseline=-3.05]\useasboundingbox ($\figScale*(-1.15,-1.25)$) rectangle ($\figScale*(2.5,1.25)$);\coordinate (v1) at ($(-90:0.8*\pLen)+\figScale*(1.05,0)$);\coordinate (v3) at ($(-3*72+36:\pLen)+(0.275,0)$);\coordinate (v5) at ($(90:0.8*\pLen)+\figScale*(1.05,0)$);\coordinate (v6) at ($(0:0.8*\pLen)+\figScale*(1.05,0)-\figScale*(0.05,0)$);\coordinate (v7) at ($(180:0.8*\pLen)+\figScale*(1.05,0)+\figScale*(0.05,0)$);\coordinate (v4) at ($(v5)-\figScale*(1.45,0)$);\coordinate (v2) at ($(v1)-\figScale*(1.45,0)$);\draw[ext] (v1)--(v2)--(v4)--(v5)--(v6)--(v1)--(v7)--(v5);\draw[ext,decorate,decoration={markings,mark=at position 0.525cm with {\arrow{stealth}}}] (v2)--(v4);\node[anchor=east] at ($(v2)!0.5!(v4)$) {{$\;\;z\!$}};\foreach \step in {-30,30} {\draw[optExt] (v1)-- ($(v1)+(-65+\step:\legLen)$);};\foreach \step in {-10,10} {\node at ($(v1)+(-65+\step:\dotDist)$) [ddot]{$$};};\node at ($(v1)+(-65:\labelDist)$) {{$$}};\node at (v1) [fullmhv] {};\draw[ext] (v2)--($(v2)+(-135:\legLen)$);\node at ($(v2)+(-135:\labelDist)$) {{$$}};\node at (v2) [fullmhv] {};\draw[ext] (v4)--($(v4)+(135:\legLen)$);\node at ($(v4)+(135:\labelDist)$) {{$$}};\node at (v4) [fullmhvBar] {};\foreach \step in {-27,27} {\draw[optExt] (v5)-- ($(v5)+(65+\step:\legLen)$);};\foreach \step in {-9,9} {\node at ($(v5)+(65+\step:\dotDist)$) [ddot]{$$};};\node[] at ($(v5)+(65:\labelDist)$) {{$\hspace{-0pt}$}};\node at (v5) [fullmhv] {};\draw[ext] (v6)--($(v6)+(0:\legLen)$);\node at ($(v6)+(0:\labelDist)$) {{$\,\,$}};\node at (v6) [fullmhvBar] {};\draw[ext] (v7)--($(v7)+(180:\legLen)$);\node at ($(v7)+(180:\labelDist)$) {{$\,\,$}};\node at (v7) [fullmhvBar] {};\node at (v6) [fullmhvBar] {};\end{tikzpicture}}
\newcommand{\schematicIntegrand}{\begin{tikzpicture}[scale=\figScale,baseline=-3.05]\useasboundingbox ($\figScale*(-1.15,-1.25)$) rectangle ($\figScale*(2.5,1.25)$);\coordinate (v1) at ($(-90:0.8*\pLen)+\figScale*(1.05,0)$);\coordinate (v3) at ($(-3*72+36:\pLen)+(0.275,0)$);\coordinate (v5) at ($(90:0.8*\pLen)+\figScale*(1.05,0)$);\coordinate (v6) at ($(0:0.8*\pLen)+\figScale*(1.05,0)-\figScale*(0.05,0)$);\coordinate (v7) at ($(180:0.8*\pLen)+\figScale*(1.05,0)+\figScale*(0.05,0)$);\coordinate (v4) at ($(v5)-\figScale*(1.45,0)$);\coordinate (v2) at ($(v1)-\figScale*(1.45,0)$);\draw[ext] (v1)--(v2)--(v4)--(v5)--(v6)--(v1)--(v7)--(v5);\foreach \step in {-30,30} {\draw[optExt] (v1)-- ($(v1)+(-65+\step:\legLen)$);};\foreach \step in {-10,10} {\node at ($(v1)+(-65+\step:\dotDist)$) [ddot]{$$};};\node at ($(v1)+(-65:\labelDist)$) {{$$}};\node at (v1) [bdot] {};\draw[ext] (v2)--($(v2)+(-135:\legLen)$);\node at ($(v2)+(-135:\labelDist)$) {{$$}};\node at (v2) [bdot] {};\draw[ext] (v4)--($(v4)+(135:\legLen)$);\node at ($(v4)+(135:\labelDist)$) {{$$}};\node at (v4) [bdot] {};\foreach \step in {-27,27} {\draw[optExt] (v5)-- ($(v5)+(65+\step:\legLen)$);};\foreach \step in {-9,9} {\node at ($(v5)+(65+\step:\dotDist)$) [ddot]{$$};};\node[] at ($(v5)+(65:\labelDist)$) {{$\hspace{-0pt}$}};\node at (v5) [bdot] {};\draw[ext] (v6)--($(v6)+(0:\legLen)$);\node at ($(v6)+(0:\labelDist)$) {{$\,\,$}};\node at (v6) [bdot] {};\draw[ext] (v7)--($(v7)+(180:\legLen)$);\node at ($(v7)+(180:\labelDist)$) {{$\,\,$}};\node at (v7) [bdot] {};\end{tikzpicture}}
\newcommand{\schematicFeynmanIntegrand}{\begin{tikzpicture}[scale=\figScale,baseline=-3.05]\useasboundingbox ($\figScale*(-0.5,-1.35)$) rectangle ($\figScale*(2.95,1.35)$);\coordinate (v1) at ($(-90:0.8*\pLen)+\figScale*(1.05,0)$);\coordinate (v3) at ($(-3*72+36:\pLen)+(0.275,0)$);\coordinate (v5) at ($(90:0.8*\pLen)+\figScale*(1.05,0)$);\coordinate (v4) at ($(v5)-\figScale*(1.05,0)$);\coordinate (v2) at ($(v1)-\figScale*(1.05,0)$);\coordinate (v7) at ($(v1)!0.5!(v5)$);\coordinate (v51) at ($(v5)+\figScale*(0.5*1,0)$);\coordinate (v52) at ($(v5)+\figScale*(0.5*2,0)$);\coordinate (v53) at ($(v5)+\figScale*(1.25,0)$);\coordinate (v11) at ($(v1)+\figScale*(0.5*1,0)$);\coordinate (v12) at ($(v1)+\figScale*(0.5*2,0)$);\coordinate (v13) at ($(v1)+\figScale*(1.25,0)$);\coordinate (v6) at ($(v7)+\figScale*(1.25,0)$);\draw[ext] (v1)--(v2)--(v4)--(v51);\draw[ext] (v53)--(v13);\draw[ext] (v11)--(v1)--(v7)--(v5);\draw[optExt] (v51)--(v53);\draw[optExt] (v11)--(v13);\draw[optExt] (v11)--($(v11)+(-90:\legLen)$);\draw[optExt] (v13)--($(v13)+(-45:\legLen)$);\node at ($(v1)+(-65:\labelDist)$) {{$$}};\node at (v1) [bdot] {};\draw[ext] (v2)--($(v2)+(-135:\legLen)$);\node at ($(v2)+(-135:\labelDist)$) {{$$}};\node at (v2) [bdot] {};\draw[ext] (v4)--($(v4)+(135:\legLen)$);\node at ($(v4)+(135:\labelDist)$) {{$$}};\node at (v4) [bdot] {};\draw[optExt] (v51)--($(v51)+(90:\legLen)$);\draw[optExt] (v53)--($(v53)+(45:\legLen)$);\node at ($(v51)!0.25!(v53)+(78.75:\dotDist)$) [ddot]{$$};\node at ($(v51)!0.5!(v53)+(67.5:\dotDist)$) [ddot]{$$};\node at ($(v51)!0.75!(v53)+(56.25:\dotDist)$) [ddot]{$$};\node at ($(v11)!0.25!(v13)+(-78.75:\dotDist)$) [ddot]{$$};\node at ($(v11)!0.5!(v13)+(-67.5:\dotDist)$) [ddot]{$$};\node at ($(v11)!0.75!(v13)+(-56.25:\dotDist)$) [ddot]{$$};\node[] at ($(v5)+(65:\labelDist)$) {{$\hspace{-0pt}$}};\node at (v5) [bdot] {};\draw[ext] (v6)--($(v6)+(0:\legLen)$);\node at ($(v6)+(0:\labelDist)$) {{$\,\,$}};\node at (v6) [bdot] {};\draw[ext] (v7)--($(v7)+(180:\legLen)$);\node at ($(v7)+(180:\labelDist)$) {{$\,\,$}};\node at (v7) [bdot] {};\node at (v51) [bdot] {};\node at (v53) [bdot] {};\node at (v11) [bdot] {};\node at (v13) [bdot] {};\end{tikzpicture}}

\begin{document}
\title{\texorpdfstring{Amplitudes at Infinity\\[-18pt]~}{Amplitudes at Infinity}}
\author{Jacob~L.~Bourjaily}
\affiliation{Niels Bohr International Academy and Discovery Center, Niels Bohr Institute,\\University of Copenhagen, Blegdamsvej 17, DK-2100, Copenhagen \O, Denmark}
\author{Enrico~Herrmann}
\affiliation{SLAC National Accelerator Laboratory, Stanford University, Stanford, CA 94039, USA}
\author{Jaroslav~Trnka}
\affiliation{Center for Quantum Mathematics and Physics (QMAP),\\Department of Physics, University of California, Davis, CA 95616, USA}

\begin{abstract}
We investigate the asymptotically large loop-momentum behavior of multi-loop amplitudes in maximally supersymmetric $(\mathcal{N}\!=\!4,8)$ quantum field theories in four dimensions. We check residue-theorem identities among color-dressed leading singularities in $\mathcal{N}\!=\!4$ supersymmetric Yang-Mills theory to demonstrate the absence of poles at infinity of all $n$-point MHV amplitudes through three loops. Considering the same test for $\mathcal{N}\!=\!8$ supergravity leads us to discover that this theory does support non-vanishing residues at infinity starting at two loops, and the degree of these poles grow {\it arbitrarily} with multiplicity. This causes a tension between simultaneously manifesting ultraviolet finiteness---which would be automatic in a representation obtained by color-kinematic duality---and gauge invariance---which would follow from unitarity-based methods.\vspace{-10pt}
\end{abstract}
\maketitle

\vspace{-20pt}\section{Introduction}\label{introduction_section}\vspace{-15pt}
Much of the remarkable recent progress in our understanding of scattering amplitudes in perturbative quantum field theory has followed from the investigation of the singularity structure of scattering amplitudes---especially when viewed at the integrand-level, as rational differential forms on the space of internal loop momenta (see {\it e.g.}~\cite{ArkaniHamed:2010kv,ArkaniHamed:2010gh,ArkaniHamed:2012nw,Bourjaily:2013mma,Bourjaily:2015jna,Bourjaily:2017wjl}). Such rational forms can be characterized by their residues, including those at infinity. Residues at finite momenta correspond to putting internal propagators on-shell, while residues at infinity often signal the need for new, irreducible kinds of contributions to scattering amplitudes. At tree-level, such boundary terms have been well studied for their role in {\it e.g.}\ the BCFW recursion relations~\mbox{\cite{Britto:2004ap,Britto:2005fq,Feng:2009ei}}.

Poles at infinity at the level of the integrand are closely related to the ultraviolet (UV) structure of integrated amplitudes. This is quite natural, as UV divergences come from singular integration regions at `infinite' loop momentum, $\ell\!\rightarrow\!\infty$. While the absence of poles at infinity implies UV finiteness of perturbative scattering amplitudes, the converse is  more subtle; the presence of poles at infinity does not necessarily imply UV divergences.

In the planar limit of $\N=4$ super Yang--Mills theory (SYM), the absence of poles at infinity is a famous consequence of the dual conformal symmetry of the theory \cite{Drummond:2008vq,Brandhuber:2008pf}. While it is not clear if any analogous symmetry exists beyond the planar limit, it has been conjectured that SYM is free of poles at infinity to all orders of perturbation theory \cite{Arkani-Hamed:2014via}. Until now, this conjecture has only been tested at relatively low orders and multiplicity---specifically, through three loops for four particles, and two loops for five particles \mbox{\cite{Arkani-Hamed:2014via,Bern:2014kca,Bern:2015ple}}. These tests were done by direct construction: by finding representations of amplitudes manifestly free of poles at infinity term-by-term. 

In this Letter, we take a different approach: we probe for the existence of non-vanishing residues at infinite loop momenta by testing whether or not residue-theorem identities are satisfied among residues at finite loop momenta alone. In particular, we make use of compact analytic formulae for color-dressed leading singularities of maximally helicity violating (`MHV') amplitudes---described in detail as an Appendix to this Letter, obtained as a generalization to the work of \cite{Arkani-Hamed:2014bca}---to directly test all such identities through three loops, thereby proving that these amplitudes are free of poles at infinity for all multiplicity. Details are included as ancillary files to this work. 

We then focus our attention on maximal $(\mathcal{N}\!=\!8)$ supergravity (`sugra'). Na\"ive power counting suggests that amplitudes in gravity behave differently due to the dimensionful coupling \cite{Bern:2011qn}. On the other hand, sugra behaves just as well as SYM at one loop for all multiplicity \cite{ArkaniHamed:2008gz}. Despite this good behavior at one loop, however, it is known that this theory supports simple poles at infinity already for four particles at three loops \cite{Bern:2014kca}. Moreover, it is expected that the degree of this pole will grow linearly with loop order, in agreement with the general expectation that four-particle amplitudes in $\mathcal{N}\!=\!8$ sugra can support UV divergences starting at seven loops \mbox{\cite{Green:2010sp,Bossard:2010bd,Elvang:2010jv,Vanhove:2010nf,Bjornsson:2010wm}}. Nevertheless, we expect there are still surprises to be found for sugra---especially at infinity \mbox{\cite{Bern:2013uka,Bern:2014sna,Herrmann:2018dja}}.

We have found that scattering amplitude integrands in $\mathcal{N}\!=\!8$ sugra behave worse at infinity than anticipated from the point of view of UV divergences of integrated amplitudes. Specifically, we demonstrate that integrands in this theory support poles at infinity \emph{starting at two loops}, and that \emph{the degree of these poles grows arbitrarily with multiplicity}. This suggests an unboundedly bad, term-wise UV behavior with increasing multiplicity, and that the theory becomes non-cut-constructible \cite{Bern:1996je} (in four dimensions) at sufficient multiplicity. To be clear, this does not imply that there are UV divergences in $\mathcal{N}\!=\!8$ at two loops; these are known to be absent from general arguments (see {\it e.g.}\ \cite{Elvang:2010jv}). Rather, this is a term-wise behavior that seems necessary within any local, gauge-invariant integrand representation of amplitudes in the theory, analogous to what was seen for four particle amplitudes in the planar limit starting at eight loops \cite{Bourjaily:2015bpz,Bourjaily:2016evz}. At the end of this Letter, we discuss color-kinematics duality \cite{Bern:2008qj,Bern:2010ue,Bern:2010yg}, and a tension between gauge-invariance and cut-constructibility.

\vspace{-20pt}\section{On-Shell Functions and (Leading) Singularities}\vspace{-15pt}
As a consequence of generalized unitarity \cite{Bern:1994cg,Bern:1994zx,Britto:2004nc}, the multi-dimensional residues of loop amplitude integrands may be computed as the product of tree-amplitudes at each vertex, summed over all the internal (on-shell) states exchanged. These objects, called {\it on-shell functions} \cite{ArkaniHamed:2012nw}, are gauge invariant and well-defined in any quantum field theory, independent of any particular representation of the amplitude. In cut-constructible theories, they represent the complete set of reference data necessary to fix perturbative amplitudes uniquely. Even if cut-constructibility were lost, the on-shell functions still carry important (yet incomplete) information about amplitudes in perturbation theory.

Historically, on-shell functions were first defined as literal residues of an off-shell loop integrand---taken on a contour which put a graph's-worth of internal propagators on-shell \cite{Bern:1994cg,Bern:1994zx,Britto:2004nc,Cachazo:2008vp}. This may be a useful picture to have in mind for this work. At $L$ loops, there are $4L$ loop-momentum degrees of freedom in four dimensions; maximal co-dimension residues---those which isolate all $4L$ degrees of freedom---are called {\it leading singularities}~\cite{Cachazo:2008vp}. They are simply algebraic functions of the external on-shell degrees of freedom (as no internal degrees of freedom remain). Next-to-maximal co-dimension residues are one-dimensional differential forms on the space of loop-momenta, etc. For example, a two-loop `heptacut'  relevant to MHV amplitudes would be:
\vspace{-8pt}\eq{\fwbox{0pt}{\hspace{-12pt}\Omega^{\mathcal{N}}\hspace{-1pt}(z)\equivR\hspace{-14pt}\egHeptacut\hspace{-5pt},\,\left\{\!\begin{array}{@{}l@{}}\ell_1(z)\!\equivR z\lambda_2\widetilde{\lambda}_1\\\ell_2(z)\!\equivR\frac{\lambda_4\big(|\ell_1\pl2\pl A|3\rangle\big)}{\ab{43}}\end{array}\right.}\label{eg_nmax_cut_os_func}\vspace{-5pt}}
where the on-shell kinematics is written in terms of spinor-helicity variables. The on-shell function (\ref{eg_nmax_cut_os_func}) can be interpreted either in $\N\!=\!4$ SYM or $\N\!=\!8$ sugra, where the `spherical' blue (white) vertices denote Bose-symmetric (color-dressed, in the case of SYM) MHV ($\overline{\text{MHV}}$) amplitudes, respectively. 

For massless theories in four dimensions, there is now a powerful framework available to represent, compute, and understand these functions: the correspondence with Grassmannian geometry described in \mbox{ref.\ \cite{ArkaniHamed:2012nw}} (see also \cite{P,Williams:2003a,ArkaniHamed:2009dn,Mason:2009qx,ArkaniHamed:2009dg,Bourjaily:2010kw,Cachazo:2017vkf}). These tools are especially powerful for ${\cal N}\!=\!4$ SYM where on-shell functions are given by the unique logarithmic differential form associated with a given cluster variety associated with the graph. The generalization to ${\cal N}\!<\!4$ SYM theory \cite{ArkaniHamed:2012nw} and gravity \cite{Herrmann:2016qea,Heslop:2016plj} is also known; but the general map between a given massless quantum field theory and a corresponding differential form is an important open question.

\vspace{-20pt}\subsection{`Residue Theorems' and Poles at Infinity}\vspace{-15pt}
Leading singularities often satisfy non-trivial relations implied by Cauchy's residue theorem \cite{cauchyTheorem}. When considering the residues of a sub-leading singularity such as (\ref{eg_nmax_cut_os_func}), the residues at {\it finite} internal momenta correspond to poles which cut an internal propagator of some vertex amplitude. Residues supported at infinite internal momenta do not have such a simple diagrammatic interpretation, and represent novel contributions to amplitudes. 

Starting from a next-to-maximal co-dimension residue (a $(4L\mi1)$-cut), let `$\partial$' denote the set of maximal co-dimension residues obtained by factorizing amplitudes at the diagram's vertices---{\it i.e.} the residues at {\it finite} internal momenta. For instance, slightly generalizing (\ref{eg_nmax_cut_os_func}) gives
\vspace{-5pt}\eq{\begin{split}\\[-15pt]\hspace{-55pt}\bigger{\partial}\hspace{-4pt}\left[\hspace{-6pt}\egGRTseed\hspace{-3pt}\right]\hspace{-3.5pt}=\hspace{-1.5pt}&\left\{\hspace{-3pt}\egGRTtermA\hspace{-4pt},\egGRTtermC,\right.\hspace{-50pt}\\[-0pt]
&\hspace{4pt}\left.\egGRTtermD,\hspace{-3.5pt}\egGRTtermB\hspace{-3pt}\right\}\!\!.\hspace{-50pt}\\[-7pt]\end{split}\label{example_grt}\vspace{00pt}}
Here, we are being schematic: the diagrams in the boundary include all those resulting from partitioning the corner vertices. The collection of leading singularities on the right hand side of (\ref{example_grt}) may or may not satisfy a `residue theorem'---as Cauchy's theorem requires that we include also any poles at infinity. However, if the on-shell functions {\it do} satisfy an identity (with $\pm1$ coefficients), then this would imply the absence of any poles at infinity. 

Thus, we may use collections of on-shell functions such as those in (\ref{example_grt}) as a diagnostic tool for testing the existence or non-existence of poles at infinity. As we will see in the next section, for $\N\!=\!8$ sugra the leading singularities in (\ref{example_grt}) do not satisfy any identity, implying the existence of poles at infinity. For $\N\!=\!4$ SYM, however, we have used such residue theorems to explicitly confirm the absence of poles at infinity for all-multiplicity MHV amplitudes through three loops.  

Recall that MHV amplitudes are given an index of $k\!=\!2$, and that an on-shell function with $n_I$ internal lines and vertex amplitudes indexed by $k_v$ has a $k$-degree of $(\sum_vk_v)\mi n_I$. There are several technical advantages which allow us to automate the test for identities such as (\ref{example_grt}) for MHV amplitudes. In particular, there exist compact analytic formulae for color-dressed cuts of MHV amplitudes which we give in the Appendix. For higher N$^{k\mi2}$MHV degree, there is no obstruction to doing a similar test, but the technical tools are less well developed and the combinatorics is more cumbersome (see \cite{Bourjaily:2016mnp}).

A complete list of on-shell functions and the seeds for residue theorems for MHV amplitudes through three loops can be found in the ancillary files for this work. Relevant, related code can be found in \cite{Bourjaily:2010wh,Bourjaily:2012gy}. Let us briefly outline the details of the checks we have done. At two loops, there are 4 distinct 8-cut vacuum/skeleton topologies, all of which support MHV coloring (for restricted arrangements of external legs); these topologies account for 6 classes of leading singularities in total (counting distinct MHV-colorings separately). There are 8 distinctly (MHV-)colored 7-cut topologies, and we have explicitly verified that each of these gives rise to an identity among leading singularities at finite momenta---thus proving the absence of poles at infinity at two loops. 

For three loops, there are 75 distinct 12-cut topologies, of which only 66 support MHV-coloring (for some restricted arrangements of external legs); these topologies account for 116 distinct classes of MHV leading singularities (again, counting distinct MHV-colorings separately). Among a total 79 distinct 11-cut vacuum/skeleton topologies, 71 admit some MHV-coloring, leading to 247 distinct potential residue-theorem identities analogous to (\ref{example_grt}) above. We have explicitly checked that all 247 identities are satisfied for generic external momenta, thus proving that these amplitudes are free of poles at infinity through three loops. 

As we have already mentioned, the collection of finite-momentum residues in (\ref{example_grt}) will fail to satisfy any identity when interpreted as on-shell functions of $\N\!=\!8$ sugra---signaling the appearance of residues at infinity. Curiously, this is the {\it only} two-loop residue theorem (among the 8) of SYM that does not hold when the on-shell functions are re-interpreted in $\N\!=\!8$ sugra. Let us now describe how this can be seen more directly, and discuss its implications for our understanding of sugra amplitudes.

\vspace{-20pt}\section{Poles at Infinity in Supergravity}\vspace{-15pt}
In \cite{Bern:2014kca}, it was found that poles at infinity are present in $\N\!=\!8$ sugra starting at three loops  for four particles, and the degree of the singularity at infinity grows (linearly) with increasing loop order. Moreover, it has been checked that there are {\it no} poles at infinity for four or five particles at two loops \cite{Arkani-Hamed:2014via,Bern:2014kca,Bern:2015ple}. As we will see, this statement fails for six or more particles already at two loops and is reflected in the failure of the collection of leading singularities on the RHS of (\ref{example_grt}) to satisfy any identity for $\N\!=\!8$ sugra. Alternatively, this can be understood by a direct computation of the heptacut (\ref{eg_nmax_cut_os_func}). In fact, an analytic expression for (\ref{eg_nmax_cut_os_func}) is not very difficult to compute (see {\it e.g.}\ \cite{Bern:2017ucb}). Indeed, the on-shell function (\ref{eg_nmax_cut_os_func}) amounts to nothing more than a BCFW shift of a one-loop leading singularity in $\N\!=\!8$ sugra,
\vspace{-5pt}\eq{\Omega^{\mathcal{N}\!=8}\hspace{-1pt}(z)=\hspace{-4pt}\schematicHeptacut\!\xrightarrow[z\to\infty]{}\frac{dz}{z^{7-n}}\,,\label{gravity_heptacut}\vspace{-4pt}}
where $n$ is the total number of external legs. Thus, this cut develops a simple pole (a logarithmic singularity) at infinity for $n\!=\!6$ particles; and the degree of the pole at infinity grows {\it arbitrarily} with the multiplicity!

The appearance of such high-degree poles at infinity has profound implications on our ability to sharply define amplitudes in this theory at the level of integrands. To see this, consider applying generalized unitarity \mbox{\cite{Bern:1994cg,Bern:1994zx,Britto:2004nc}}, starting with a complete basis of rational integrands consistent with a bounded polynomial degree of loop-dependence in integrand numerators. (See {\it e.g.}\ refs.\ \mbox{\cite{Passarino:1978jh,Ossola:2006us,Bourjaily:2017wjl,Ben-Israel:2018ckc}} for how bases of integrands may be constructed.) Using only irreducible integrands (those with at most $4L$ propagators in four dimensions), the cut (\ref{gravity_heptacut}) would be matched by an integrand of the form
\vspace{-5pt}\eq{\hspace{-20pt}\schematicIntegrand\quad\text{with}\quad N(\ell)\!\sim\!(\ell\!\cdot\!Q)^{n-4}\,,\hspace{-20pt}\label{nmax_local_diagram_scaling}\vspace{-5pt}}
where $N(\ell)$ represents the loop-dependent numerator degrees of freedom of integrands with this topology in the basis, and $Q$ denotes some external four-momenta. The exact form of these numerators is not important for us at present. However, the general form of the scaling in (\ref{nmax_local_diagram_scaling}) has two immediate, and important consequences. 

Firstly, if we consider the basis of integrands consistent with the power counting implied by (\ref{nmax_local_diagram_scaling}), it is clear that for sufficiently large $n$ these integrands will scale worse than pure Yang-Mills \cite{Feng:2012bm}. The methods described by OPP \cite{Ossola:2006us} can be used to fix the coefficients of the integrals in this basis {\it from the Feynman expansion}; but in general, these coefficients will not be cut-constructible. This would seem to suggest that $\N\!=\!8$ is not a cut-constructible theory in general, although this conclusion may be premature. 

Another consequence of the scaling in (\ref{nmax_local_diagram_scaling}) is that, for sufficiently large multiplicity, these integrands will be UV divergent(!). This is despite the fact that two-loop amplitudes in $\N\!=\!8$ are UV finite for all multiplicity \cite{Elvang:2010jv}. The fact that the asymptotic behavior of the cut (\ref{gravity_heptacut}) does not in fact signal any UV divergences in the theory is easy to see from color-kinematic duality \mbox{\cite{Bern:2008qj,Bern:2010ue,Bern:2010yg}}---provided, that is, that such a representation exists for all two loop amplitude integrands in $\N\!=\!8$ sugra. But such representations always exist for {\it cuts} such as (\ref{gravity_heptacut}).

Using color-kinematic satisfying representations of the tree amplitudes in (\ref{gravity_heptacut}) (which always exist \mbox{\cite{Bern:2010yg,BjerrumBohr:2010yc,Bjerrum-Bohr:2016axv}}), we would find a representation of this cut in sugra in terms of cubic graphs with $(n\pl2)$ vertices and $(n\pl3)$ propagators---schematically, an integrand of the form
\vspace{-6pt}\eq{\fwbox{0pt}{\hspace{-18pt}\schematicFeynmanIntegrand\hspace{-2pt}\sim\hspace{-2pt}\int\!\!\frac{(d^4\hspace{-1pt}\ell)^2}{(\ell^2)^7}\frac{(\ell\!\cdot\!Q)^{2n-8}}{(\ell^2)^{n-4}}\!\stackrel{\text{cut}}{\longrightarrow}\!\!\int\!\!\frac{dz}{z^3}\frac{z^{2n-8}}{z^{n-4}}\,,\!}\label{schematic_accounting_of_degree}\vspace{-6pt}}
where, again, $Q$ is either an external momentum or polarization and both $(\ell\cdot Q)$ and $\ell^2$ scale linearly with $z$. In (\ref{schematic_accounting_of_degree}), we have taken into account $\mathcal{N}\!=\!8$ supersymmetry by factoring-out eight powers of loop-momentum independent kinematic invariants. Importantly, it is easy to see that the integrand (\ref{schematic_accounting_of_degree}) is UV finite (by na\"ive power counting) for any $n$. The resolution of the apparent paradox is that integrands such as (\ref{schematic_accounting_of_degree}) are generically {\it reducible} (in the sense of \cite{Passarino:1978jh}) in any integrand basis with fixed spacetime dimension; and the expansion of (\ref{schematic_accounting_of_degree}) into irreducible integrands such as (\ref{nmax_local_diagram_scaling}) represents a UV finite integral in terms of many UV divergent pieces. It is worth noting that integrands obtained by squaring individual terms from SYM would necessarily not be gauge invariant. 

\vspace{-20pt}\section{Conclusions and Outlook}\vspace{-15pt}
In this Letter, we have shown that (MHV) amplitudes in $\N\!=\!4$ SYM are free of poles at infinity through three loops by directly testing `residue theorem' identities involving only residues at finite loop momenta. This directly tests the conjecture of \mbox{ref.\ \cite{Arkani-Hamed:2014via}}, strengthening the suspicion that perturbative amplitudes in full SYM may have additional, still undiscovered symmetries. 

Applying the same tests to $\N\!=\!8$ sugra, however, leads us to the discovery that amplitude integrands in this theory behave arbitrarily worse than SYM for large multiplicity even at two loops. While the existence of poles at infinity does not immediately impede our ability to construct representative integrands for a theory, it may signal that such integrands cannot be defined using purely gauge-invariant data. As we have seen for $\N\!=\!8$ sugra, simply `matching all cuts' cannot suffice, because there are polynomial degrees of freedom of arbitrarily high degree which cannot be fixed by cuts. Moreover, an integrand basis capable of matching these degrees of freedom in the numerator would necessarily have arbitrarily bad UV behavior term-by-term. 

Despite this term-wise UV behavior, we have seen that color-kinematics duality implies this is an artifact of an integrand basis consisting of only irreducible topologies. This is consistent with the general claim that $\N\!=\!8$ sugra is UV-finite for all multiplicity at this order of perturbation \cite{Elvang:2010jv}. Using generalized gauge freedom \cite{Bern:2010yg} allows us to represent the ingredients of the double-copy in a multitude of ways; and it is {\it a priori} unclear if different double-copy formulae are equivalent at  the integrand level (although the integrated amplitudes must certainly be equal). Said another way, it is not clear if BCJ duality makes sense in a {\it strictly} gauge-invariant way. This is of course a problem of the Feynman representation of amplitude integrands, which are known not to be gauge invariant outside of special classes of theories.

On the other hand the recent analysis \cite{Herrmann:2018dja} showed that certain anticipated poles at infinity are surprisingly absent in sugra requiring cancelations between terms in the integrand basis. This suggests there exist additional conditions---likely related to the behavior at infinite momenta---which can make the sugra integrand well-defined. The existence of a unique, gauge-invariant integrand seems like a necessary first step in the search for a potential geometric definition of the amplitude as was the case for planar $\N\!=\!4$ SYM \cite{Arkani-Hamed:2013jha}; but we must leave such questions to future work. 
\nopagebreak

\vspace{-20pt}\section{Acknowledgments}\vspace{-15pt}
This project has been supported by an ERC Starting Grant \mbox{(No.\ 757978)} and a grant from the Villum Fonden (JLB); and by a grant from the Simons Foundation (341344, LA) (JLB).
The research of J.T. is supported in part by U.S. Department of Energy grant DE-SC0009999 and by the funds of University of California. E.H. is supported by the U.S. Department of Energy under contract DE-AC02-76SF00515.

\appendix
\vspace{-20pt}\section{Color-Dressed On-Shell Functions in SYM}\vspace{-15pt}
In order to test the potential `residue theorems' among on-shell functions such as those in (\ref{example_grt}) for $\N\!=\!4$ SYM beyond the planar limit, we needed efficient representations of all such functions---including their full color dressing.

We start by decomposing the Bose-symmetric, color-dressed vertex (tree-)amplitudes of a leading singularity according to Del Duca, Dixon and Maltoni (DDM) \cite{DelDuca:1999rs}:
\vspace{-4pt}\eq{\mathcal{A}^{YM}_n({\color{anchorLeg}\alpha},{\color{unord}\hspace{0pt}A},{\color{anchorLeg}\beta})=\sum_{\fwbox{25pt}{{\color{ord}\vec{a}}\!\in\!\mathfrak{S}\hspace{-1pt}(\hspace{-1pt}{\color{unord}\hspace{-0.0pt}A}\hspace{-0.75pt})}}\ddmColor{\alpha}{a}{\beta}A^{YM}_{n}({\color{anchorLeg}\alpha},{\color{ord}\vec{a}},{\color{anchorLeg}\beta})\,,\label{ddmification}\vspace{-6pt}}
where the anchored legs $\{{\color{anchorLeg}\alpha},{\color{anchorLeg}\beta}\}$ are arbitrary and the sum is over the $(n\mi2)!$ permutations ${\color{ord}\vec{a}}\equivL\!\{{\color{ord}a_1},{\color{ord}\ldots},{\color{ord}a_{\text{-}1}}\}$ of the set $\hspace{-1pt}A\!\equivR\![n]\backslash\{{\color{anchorLeg}\alpha},{\color{anchorLeg}\beta}\}$. (Notice that we have used `$\mi1$' to denote the final entry of an ordered list.) For each ordered set ${\color{ord}\vec{a}}$, $A^{YM}_{n}({\color{anchorLeg}\alpha},{\color{ord}\vec{a}},{\color{anchorLeg}\beta})$ represents the color-stripped, (but \mbox{color-)}ordered partial amplitude, and $\ddmColor{\alpha}{a}{\beta}$ denotes the color structure
\vspace{-5pt}\eq{\begin{split}\ddmColor{\alpha}{a}{\beta}&\equivR\hspace{-3pt}\sum_{e_i}f^{{\color{anchorLeg}\alpha},{\color{ord}a_1},e_1}f^{e_1,{\color{ord}a_2},e_2}\cdots f^{e_{-1},{\color{ord}a_{-1}},{\color{anchorLeg}\beta}} \text{}\\[-14pt]
&\equivL\hspace{4pt}\ddmFgraphExpanded\equivL\ddmFgraphContracted\,,\\[-13pt]
\end{split}\label{ddm_color_factor_defined}}
where the (adjoint representation) color matrix of the $a$th supermultiplet is given by the structure constants of some Lie algebra, $(T^a)_{bc}\equivR\! f^{abc}$. Notice our graphical notation for  $\ddmColor{\alpha}{a}{\beta}$ defined in (\ref{ddm_color_factor_defined}) allows for ${\color{ord}\vec{a}}\!=\!\{\}$; in this case, (\ref{ddm_color_factor_defined}) should be understood as giving the color factor $\delta^{}_{{\color{anchorLeg}\alpha}{\color{anchorLeg}\beta}}$. Although the partial-ordered amplitudes are always cyclically symmetric, the color structures $\ddmColor{\alpha}{a}{\beta}$ never are (unless ${\color{ord}\vec{a}}$ is a single leg); thus, although in (\ref{ddm_color_factor_defined}) we are using a single vertex to denote this structure, the differentiation among legs is necessary to define the RHS.

Using the DDM decomposition, we may represent any Bose-symmetric, color dressed cut of $\N\!=\!4$ SYM in terms of color factors built from (\ref{ddm_color_factor_defined}) and (color-)ordered partial amplitudes. Using flat discs to denote color-stripped partial amplitudes, we could decompose, for example,
\vspace{-6.5pt}\eq{\raisebox{29pt}{\fwbox{0pt}{\hspace{-5pt}\egHexaboxFull\hspace{-8pt}{=\!\!\displaystyle\sum_{\hspace{-13.5pt}\substack{\vec{a}\in\mathfrak{S}\hspace{-1pt}(\hspace{-1pt}A\hspace{-0.75pt})\\
(\hspace{-1pt}\vec{b}_1\hspace{-1pt}\cup\hspace{-0.5pt}\vec{b}_2\hspace{-1pt})\in\mathfrak{S}\hspace{-1pt}(\hspace{-1pt}B\hspace{-0.75pt})\\
(\hspace{-1pt}\vec{c}_1\hspace{-1pt}\cup\hspace{-0.5pt}\vec{c}_2\hspace{-1pt})\in\mathfrak{S}\hspace{-1pt}(\hspace{-1pt}C\hspace{-0.75pt})}\hspace{-13.5pt}}}\hspace{-8pt}\egHexaboxColour\hspace{-14pt}\times\hspace{-10pt}\egHexaboxOrdered\fwboxL{0pt}{\hspace{-10pt}.}}}\label{eg_color_dressed_on_shell_function_in_sym}\vspace{-5pt}}
The color structures built from graphs of  $\ddmColor{\alpha}{a}{\beta}$'s can easily be expanded into a trace basis once a particular Lie algebra ({\it e.g.}\ $SU(n)$) is specified. The kinematic factors appearing in (\ref{eg_color_dressed_on_shell_function_in_sym}) turn out to be quite interesting.

\vspace{-20pt}\subsection{(Color-)Ordered On-Shell Functions in SYM}\vspace{-15pt}
Diagrams similar to those appearing in (\ref{eg_color_dressed_on_shell_function_in_sym}) for MHV amplitudes were studied in \mbox{ref.\ \cite{Arkani-Hamed:2014bca}}---specifically, the case where leading singularities were constructed purely from trivalent vertices. This simplifies matters considerably, as even without color factors, three particle amplitudes in SYM are permutation invariant up to a sign. 

As was found in \mbox{ref.\ \cite{Arkani-Hamed:2014bca}}, there are two differently-elegant descriptions for the kinematic-dependent functions associated with diagrams such as those appearing in (\ref{eg_color_dressed_on_shell_function_in_sym}). As was the case for trivalent diagrams, each ordered MHV amplitude at a vertex depends cyclically and holomorphically on some number of $\lambda$-spinors. Whether these are external legs {\it directly} entering the vertex, or via some chain of $\overline{\text{MHV}}$ vertices makes no difference. Thus, these functions are specified by a list of (individually cyclically-invariant) lists of external legs attached to each MHV amplitude in the diagram, $\Gamma(\vec{a},\vec{b},\ldots,\vec{c})$. (The sizes of these lists are not fixed; but the sum of their lengths is related to the number of external legs of the diagram.)

Let us start with the conceptually simplest formula for a function labeled in this way. For a given permutation (ordered list of) of external particle labels $\vec{\sigma}$, let $\text{PT}(\vec{\sigma})\equivR1/(\ab{\sigma_1\sigma_2}\cdots\ab{\sigma_{{-}1}\sigma_{1}})$; then the on-shell function labeled by the lists $\{\vec{a},\ldots,\vec{c}\}$ may be expressed
\vspace{-2pt}\eq{\Gamma\big(\vec{a},\ldots,\vec{c}\big)=\sum_{\fwboxL{30pt}{\hspace{-5pt}\sigma\!\in\!\vec{a}\!\shuffle\!\vec{b}\!\shuffle\!\cdots\!\shuffle\!\vec{c}}}\text{PT}(\vec{\sigma})\,,\label{shuffle_sum_formulae}\vspace{-2pt}}
where the sum is over all `cyclic' shuffles $\sigma$---permutations of the set $[n]$ consistent with the {\it cyclic} ordering of each of $\{\vec{a},\vec{b},\ldots,\vec{c}\}$. (That is, all permutations $\sigma$ of the particle labels for which the intersection of $\sigma$ with each set represents a cyclic permutation of that set.) For example,~\\[-10pt]
\vspace{-7pt}\eq{\egHexaboxOrdered\hspace{-5pt}=\sum_{\hspace{-0pt}\widehat{a}\in\vec{a}\shuffle(\vec{c}_2|\delta|\vec{b_2})\hspace{-30pt}}\text{PT}(\alpha,\widehat{a},\beta,\vec{b}_1,\gamma,\vec{c}_1)\,,\vspace{-5pt}}
where `$|$' denotes the concatenation of ordered lists. 

Let us now describe the second, more compact (but more involved) representation of the same function. For each ordered vertex of the diagram, $\vec{a}$, define the vector
\vspace{-2pt}\eq{\begin{split}\vec{\psi}(\vec{a})\!&\equivR\!\!\{\psi_1(\vec{a}),\ldots,\psi_{|\vec{a}|-2}(\vec{a})\},\,\,\text{with components}\\
\psi_j(\vec{a})\!&\equivR\!\!\Big(\lambda_{a_j}\frac{\ab{a_{j+1}a_{j+2}}}{\ab{a_{j}a_{j+1}}}\pl\lambda_{a_{j+1}}\frac{\ab{a_{j+2}a_{j}}}{\ab{a_{j}a_{j+1}}}\pl\lambda_{a_{j+2}}\Big)\,.\\[-6pt]\end{split}\label{psi_columns_defined}\vspace{-5pt}}
A vertex involving $|\vec{a}|$ legs is thus associated with a vector $\vec{\psi}(\vec{a})$ of length $|\vec{a}|\mi2$. From these vectors, we construct the matrix of (concatenated) derivatives
\vspace{-2pt}\eq{\Psi\equivR\partial_{\lambda}\Big(\vec{\psi}(\vec{a})\Big|\vec{\psi}({\vec{b}})\Big|\cdots\Big|\vec{\psi}({\vec{c}})\Big)\,\vspace{-2pt}}
where $\partial_{\lambda}$ represents the derivative of each column (\ref{psi_columns_defined}) with respect to the $n$ external $\lambda$'s. Thus, $\Psi$ is an $n\!\times\!(n\mi2)$ matrix of rank $n\mi2$. Deleting rows $\{j,k\}$ will result in a matrix of full rank, denoted $\Psi_{j,k}$. One can check that
\vspace{-2pt}\eq{\hspace{-10pt}\mathfrak{J}\equivR\!\!\det\!\big(\!\Psi_{j,k}\!\big)\big[\ab{a_1a_2}\ab{b_1b_2}\cdots\ab{c_1c_2}\big]/\ab{j\,k}\,\hspace{-10pt}\label{psi_det}\vspace{-2pt}}
is both independent of the choice $\{j,k\}$, and cyclically invariant with respect to each list $\{\vec{a},\vec{b},\ldots,\vec{c}\}$. (Notice that the cyclic-invariance of each list was broken in (\ref{psi_columns_defined}) (by the choice to include only the first $|\vec{a}|\mi2$ components); the asymmetry of this choice is reflected in the angle brackets in the numerator of (\ref{psi_det}).) In terms of the matrix $\Psi$, a more compact alternative to (\ref{shuffle_sum_formulae}) is simply:
\vspace{-2pt}\eq{\hspace{-20pt}\Gamma\big(\vec{a},\cdots,\vec{c}\big)=\Big[\text{PT}\big(\vec{a}\big)\text{PT}\big(\vec{b}\big)\cdots\text{PT}\big(\vec{c}\big)\Big]\!\times\!\mathfrak{J}^2\,.\hspace{-20pt}\vspace{-2pt}}
This formula, while compact and elegant, obscures the simple fact that every MHV on-shell diagram (for SYM) has only simple poles which in turn is manifest from (\ref{shuffle_sum_formulae}).

\vspace{-12pt}
\providecommand{\href}[2]{#2}\begingroup\raggedright\endgroup

\end{document}